# Low-temperature ballistic transport in nanoscale epitaxial graphene cross junctions

S. Weingart, C. Bock, and U. Kunze

Werkstoffe und Nanoelektronik, Ruhr-Universität Bochum, D-44780 Bochum, Germany

F. Speck, Th. Seyller, and L. Ley

Technische Physik, Friedrich-Alexander-Universität Erlangen-Nürnberg,

D-91058 Erlangen, Germany

We report on the observation of inertial-ballistic transport in nanoscale cross junctions fabricated from epitaxial graphene grown on SiC(0001). Ballistic transport is indicated by a negative bend resistance of $R_{12,43} \approx -170\ \Omega$, which is measured in a non-local, four-terminal configuration at 4.2 K and which vanishes as the temperature is increased above 80 K.

Graphene, a two-dimensional, honeycomb lattice of carbon atoms, has unique electronic properties such as a high charge carrier mobility which only weakly depends on the temperature[1] and charge carriers which can be described as massless relativistic quasi-particles[2] an account of their linear energy dispersion relation[3] and the chiral nature of their wave function. Since, even at room temperature, the ballistic mean free path of charge carriers in exfoliated[4] and epitaxial[5] graphene achieves values of several 100 nm, ballistic devices based on graphene are of great interest not only for fundamental studies but also for future applications. Nevertheless, up to now only a few experiments show signatures of ballistic transport.[6,7] In exfoliated graphene at low temperatures phase coherent transport is shown with superconducting contacts[6] and in Aharonov-Bohm graphene rings.[7] For novel devices large-scale graphene films on an insulating substrate are required. This is provided by epitaxial growth of graphene on SiC.[8,9] Significant improvements of the growth technique for graphene on SiC(0001)[10,11] accompanied by improved morphology and electronic properties[12] allows us to observe inertial-ballistic transport on epitaxial graphene. Here we report on the observation of negative bend resistance found in a nanoscale orthogonal graphene cross junction.

Starting material for the preparation of Hall-bars and nanoscale cross junctions is a monolayer graphene epitaxially grown on the Si-face of a semi-insulating 6H-SiC substrate. The graphene film is formed by annealing the SiC substrate at $T = 1650$ °C in 1 bar Ar. Details of the preparation process of the graphene film are described elsewhere.[10] Fig. 1(a) shows a top-view atomic force micrograph of the graphene film. Characteristic terraces with a width of about $1 - 2$ µm and a step height of $2 - 4$ nm are formed in the substrate. As shown by low-energy electron microscopy the terraces are homogeneously covered with a graphene monolayer.[10] Additionally, small stripes of bilayer graphene are formed at the step edges. Hall-bar structures, 700 nm wide and 3 µm long, are employed to determine the charge carrier mobility $\mu$, electron density $n_s$, and the mean free path $l_e$. To study ballistic effects in bend resistance geometry orthogonal cross junctions with four identical leads of 50 nm (80 nm) width and 400 nm length are realized [Fig. 1(b)]. The devices are prepared by using a mix-and-match technique. In the first step nanoscale cross junctions and mesoscopic Hall-bars are defined by low-energy electron beam lithography in a high-resolution calixarene resist. Subsequently, large area leads and contacts are specified by conventional lithography in a standard UV resist.[13] Finally, the combined resist pattern is transferred into the graphene layer in a single step etch process. Low-damage dry etching is performed by inductively coupled oxygen plasma at a RF bias of −19 V. The ohmic contacts of current leads and voltage probes are formed by Ti/Au pads which are prepared by conventional lithography, metal evaporation and lift-off.

The electron density $n_s$ and mobility $\mu$ are derived from Shubnikov-de Haas oscillations and temperature dependent Hall-effect measurements. The results indicate that the graphene film is highly n-doped with a constant electron density of $n_s \approx 2.6 \cdot 10^{12}$ cm$^{-2}$ in the temperature interval 1.4 K $\leq T \leq$ 300 K.[12] At temperatures below 50 K the electron mobility amounts $\mu \approx 3080$ cm$^2$ (Vs)$^{-1}$ and, as temperature is increased (50 K $\geq T \geq$ 300 K), linearly decreases to 1640 cm$^2$ (Vs)$^{-1}$.[12] The electron mean free path is estimated from a semi-classical model to $l_e = (\hbar/e) \cdot \mu \cdot (\pi n_s)^{1/2}$,[14] where $\hbar$ is the Planck constant divided by $2\pi$ and $e$ the elementary charge. The result is a mean free path of 58, 55, and 31 nm at T = 4.2, 80, and 300 K, respectively. For low temperatures the electron mean free path exceeds the width of the cross junction ($b$ = 50 nm). Under this condition we expect that the transfer characteristic of the device is dominated by inertial-ballistic effects.[15] As temperature is increased a transition from the ballistic into the diffusive transport regime should be observable in the four terminal transfer characteristic.

The cross junctions are studied in bend resistance geometry by dc transfer measurements at different temperatures. An input voltage $V_{12} = V_1 - V_2$ is applied in push-pull configuration ($V_1 = -V_2$) between the orthogonal leads "1" and "2" and induces a current $I_1$. The potential difference $V_{43} = V_4 - V_3$ is measured between the adjacent voltage probes "4" and "3". We use an Agilent Precision Semiconductor Parameter Analyzer 4156B which enables us to simultaneously detect $I_1$ and $V_{43}$. The bend resistance is defined by $R_{12,34} = V_{43}/I_1$. In the diffusive transport regime the electrons drift along the electric field lines. By applying an input voltage $V_{12} > 0$ ($V_{12} < 0$) a positive (negative) potential difference $V_{43}$ is induced and the bend resistance is positive. In the ballistic regime the mean free path of the electrons exceeds the width of the junction ($l_e > b$) and the electrons are injected from the negatively biased current lead into the opposite voltage probe. Consequently, for $V_{12} > 0$ ($V_{12} < 0$) the voltage probe "4" ("3") is negatively charged and the transfer voltage $V_{43}$ is negative (positive). Therefore, in the ballistic transport regime a negative bend resistance $R_{12,43}$ is observed. The transfer characteristic of the bend resistance is linear.[16]

Fig. 2 shows the bend resistance characteristics of a graphene cross junction at different temperatures. Multiple cooling cycles are performed to ensure the reproducibility of the observed features. At low temperatures ($T = 4.2$ K) the negative bend resistance of $R_{12,43} \approx -170\ \Omega$ indicates the ballistic regime.

For $T = 77$ K a negative bend resistance is still observed, even though the value $R_{12,43} \approx -8\ \Omega$ is considerably smaller. Since the mean free path does not change significantly between 4.2 and 80 K in the Hall-bar devices a nearly constant $R_{12,43}$ would have been expected below 80 K. The variation of $R_{12,43}$ in the cross junction indicates that additional scattering plays an important role which we attribute to the device boundaries. At even higher temperature the electron-phonon scattering rate increases and the electron mean free path is reduced below the relevant device dimensions. Consequently, for $T = 300$ K a change of sign is observed in $R_{12,43}$ and the transfer characteristic shows the well-known linear ohmic behaviour of the diffusive transport regime. For cross junctions with leads of width $b = 80$ nm no negative bend resistance is observed, which is in accordance with the estimated mean free path ($l_e < 80$ nm).

Due to the geometrical symmetry of a cross junction the bend resistance characteristic of an ideal device is expected to be indepent of the specific pair of leads used for current injection. In our device we found a similar transfer characteristic for $V_{43}(I_1)$ and $V_{21}(I_3)$, whereas the transfer characteristics of $V_{32}(I_4)$ [Fig. 3] and $V_{14}(I_2)$ show a strongly reduced ballistic signal at $T = 4.2$ K, and at $T = 77$ K the positive bend resistance indicates diffusive transport. Compared to modulation-doped high-mobility GaAs/AlGaAs[17] and Si/SiGe[18] heterostructures which are known to show negative bend resistance, two striking differencies may explain the observed asymmetry in epitaxial graphene: (1) In graphene the two dimensional electron gas is directly located on top of the substrate and is not protected by any cap layer. Thus, in contrast to the above mentioned heterostructures the electron transport is strongly affected by scattering at impurities and adsorbates on top of the sample. (2) Transitions from monolayer graphene to bi- or trilayer graphene at the step edges may cause additional scattering since the cross junction straddles two or more terraces. A typical arrangement of step edges running through a cross junction can be found in the scanning electron micrograph of a similar test structure [Fig. 1(b)]. In order to study the influence of step edges on the ballistic transport in detail further work on cross junctions positioned on a terrace with different displacements relative to step edges is necessary.

In summary, we demonstrate low-temperture ballistic transport in a nanoscale epitaxial graphene cross junction. The advanced morphology of graphene films on SiC(0001) grown in an Ar atmosphere yields in a electron mean free path of $l_e = 58$ nm at $T = 4.2$ K. The mean free path exceeds the device dimensions for a 50 nm wide othogonal cross junction and a negative bend resistance of $R_{12,43} = -170$ Ω observed at $T = 4.2$ K indicates ballistic transport. With increasing temperature a transition from the ballistic to the diffusive transport regime is signalled by a change of sign in $R_{12,43}$. Asymmetries in the bend resistance measured from wiring different pairs as current injecting leads indicate scattering at substrate step edges.

This work was supported by the Ruhr University Research School funded by the DFG in the Framework of the Excellence Initiative. Work in Erlangen was supported by the DFG under Grant SE1087/5-1 and by the excellence cluster "Engineering of Advanced Materials" (www.eam.uni-erlangen.de) at the Friedrich-Alexander-Universität Erlangen-Nürnberg.

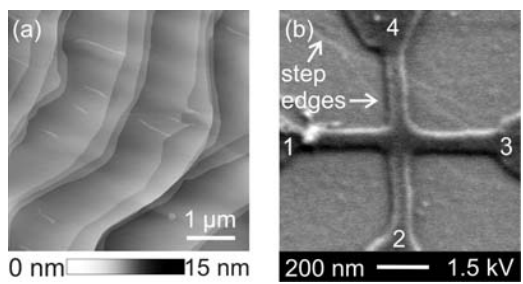

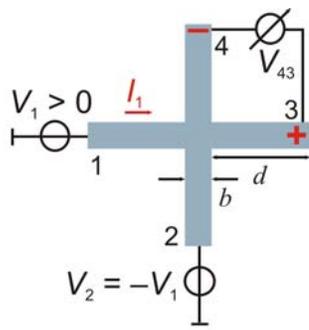

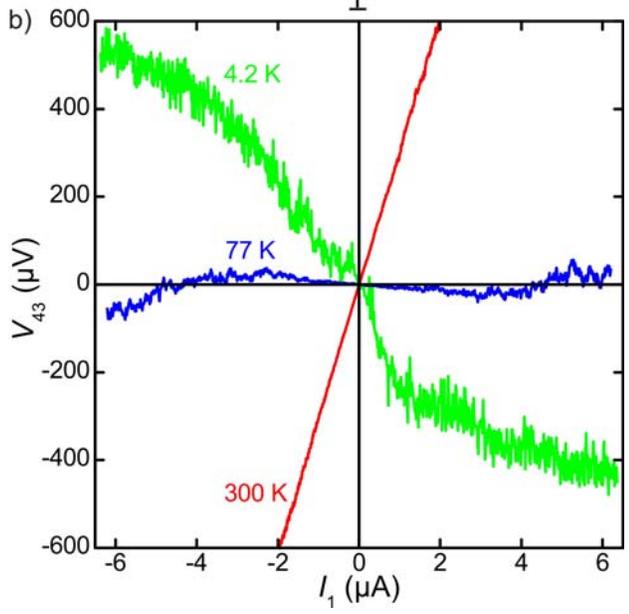

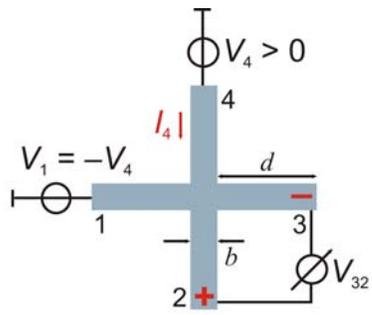
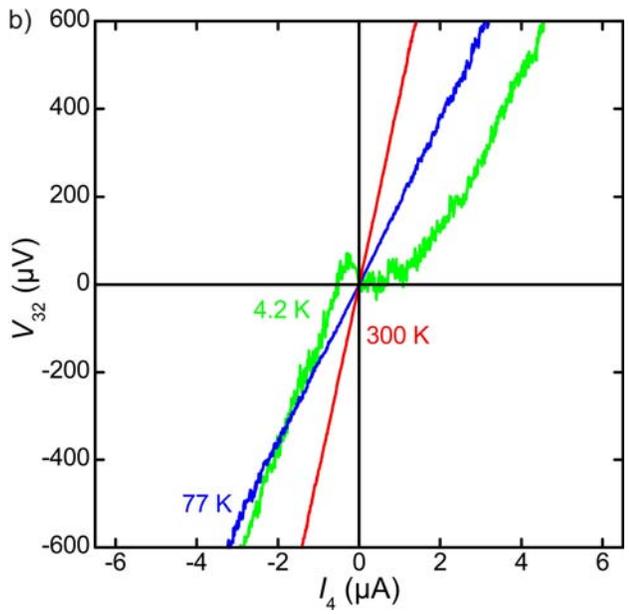

**Fig. 1.** (a) Atomic force micrograph of epitaxial graphene on a Si-faced SiC surface grown in Ar. (b) Oblique-view scanning electron micrograph of a test structure. The cross junction ($b \approx 50$ nm) is patterned from a single layer of epitaxial graphene (dark). The marked lines correspond to step edges in the substrate.

**Fig. 2.** (a) Wiring of current leads and voltage probes. (b) Bend voltage $V_{43}$ as a function of input current $I_1$ of an orthogonal cross junction with four identical leads ($b = 50$ nm, $d = 400$ nm) at different temperatures. For $T = 4.2$ K the negative bend resistance demonstrates ballistic transport.

**Fig. 3.** (a) Wiring of current leads and voltage probes turned by 90° with respect to those used in Fig. 2. (b) Bend voltage $V_{32}$ vs $I_4$ of the same cross junction as used in Fig. 2 at $T = 4.2$ K, $T = 77$ K and $T = 300$ K.